\documentstyle[wrapfig,epsf,epsfig,rotating]{elsart}
\begin{document}
\begin{frontmatter}
\title{{\sf RAMPEX} - a new spin experiment}
\author{Yu. Arestov}
\address{ Institute for High Energy Physics,
142284 Protvino, Moscow region, Russia; 
{\rm E-mail}: arestov@mx.ihep.su \\[2mm]
{\rm (presented for HELION97, 20-24 Jan 1997, Kobe, Japan)}
}
\begin{abstract}
{\sf RAMPEX} - Russian-AMerican Polarization EXperiment-
is dedicated to  studies of one-spin asymmetries which have twist-3 and 
also twist-2 origin, in hard and semi-hard inclusive processes on the
polarized propane-diol target. A special consideration has been given 
for the prospects of using polarized $^3$He target. The studies will be 
performed at the Serpukhov accelerator at 70 GeV/c ($p$ beam) and 40 GeV/c 
($\pi^-$ beam). 

\end{abstract}

\end{frontmatter}

\section {Introduction.}
   {\sf RAMPEX} -- Russian-AMerican Polarization EXperiment--
presents a program of studies of  one-spin effects in hadron processes.
A significant part of the investigations will be performed for the first time.
This program includes $pp_{\uparrow}$ interactions at 70 GeV/c, and as an
option $\pi^{-}p_{\uparrow}$ interactions at 40 GeV/c, at the Serpukhov
accelerator. Conceptually {\sf RAMPEX} will try to form a new approach in 
interpreting experimental results on one-spin asymmetries in hadron processes.

   Quantum chromodynamics as a model of strong interactions is commonly used
in interpretation of experimental data. The most successful  descriptions 
hold for
those effects which correspond to the leading twist-2 contributions. These 
include collider data on hard photon and jet production 
at large $p_T$, on lepton pair production with large $M$ etc. 

   The higher-twist contributions are less familiar to most of physicists.
The twist-3 phenomena relate to the one-spin processes, such as 
hard and semi-hard hadron production 
on polarized protons in reactions 
\begin{equation}
   pp_{\uparrow} \rightarrow h+X  \label{h}
\end{equation}
at high
initial energies. Though one-spin asymmetries were studied experimentally
at various energies (see for example review [1]), no special attention
was paid to the twist-3 origin of the one-spin asymmetries.

   There are also subtle twist-2 effects  in the double inclusive processes
\begin{equation}
    pp_{\uparrow} \rightarrow h_{1}+h_{2}+X  \label{hh}
\end{equation}
which have not yet been studied.

   The spin theoretical community is highly interested in the new results
concerning these twist-3 and twist-2 hadron processes (\ref{h},\ref{hh})
(see review [2]).
The forthcoming {\sf RAMPEX} stimulates them to formulate new concepts for 
one-spin asymmetries.

   Analysis of  the whole set of experimental measurements which are 
available between  beam momenta of 6 and 200 GeV/c (for refs. see [1]), 
results in the following conclusions:
\begin{itemize}
   \item [--] at any initial energy under study, the one-spin
                  asymmetry can reach sizeable values;
   \item [--] a serious enigma is the zero values of $A_{N}(p_{T})$ in
              a very wide $p_{T}$-interval in E704 experiment at 200 GeV/c;
   \item [--] the experimental measurements are rather mosaic, and mostly
                     they cannot be compared;
   \item [--]  a new experiment is highly desirable in which 
               a complex physical program  can be performed 
               on $A_{N}(x_{F},p_{T})$ for different particles
               and in different kinematical regions.
\end{itemize}

\section { Physical motivation of {\sf RAMPEX}.}  
Normally the experimental community is 
familiar with the polarized parton densities $g_{1}(x)$  and $g_{2}(x)$ 
probed in DIS. However the intrinsic nucleon structure is described also 
by other functions, of twist-2 and twist-3 in particular (see for 
example [3]).

In {\sf RAMPEX} we shall measure the  twist-3 
asymmetries in single hadron production in (\ref{h})
and we shall try to discover the more subtle twist-2 correlations in
two-particle production processes in (\ref{hh}).
In both cases $h$'s denote $\pi ,~K$,... As is commonly believed, 
the one-spin asymmetries in hard and
semi-hard hadron production processes can be used,
in appropriate phenomenology, to obtain
information on new spin-dependent quark distributions 
$h_{L}(x)$ and $h_{1}(x)$, of twist-3 and twist-2 respectively. They 
are both chiral-odd distribution functions.

 The relevant theoretical problems were 
discussed at the {\sf RAMPEX} Round Table at SPIN96 [2].

   The decaying $\Lambda$-hyperons are good self-analysing polarization
tools. The measurment of the final-state $\Lambda$ polarization in the 
reaction
   $$  p + p_{\uparrow} \rightarrow \Lambda_{\uparrow} + X $$
will allow to study the proton--$\Lambda$ spin correlations. We shall 
also perform first polarization asymmetry measurements in the production
processes of the resonanses $K^{\ast}_{890}$ and $\phi$ containing 
$s$ quarks.

 \section { General layout.}
The full version of the experimental setup includes two arms (fig. 1).
One arm consists of the magnet spectrometer, two \v Cerenkov counters 
{\sf \v C1, \v C2}
to identify charged particles, an electromagnetic calorimeter {\sf EC1} 
and a hadron calorimeter {\sf HC}.  The magnet spectrometer consists of
the magnet {\sf M} and five proportional chambers {\sf PC1-PC5}.
In fig. 1 this arm makes an angle of $9^{\circ}$ with the beam line 
corresponding to $90^{\circ}$ in cms. 
This arm will be also rotated to a smaller angle close to $0^{\circ}$ 
to detect particles with large $x_F$ and to a larger angle
to detect particles with negative $x_F$. Numerical estimations of 
acceptances and efficiences show that the angle values near 80 and 300 
mrad are optimal for these measurements.
   The second arm of the setup consists only of the fine-granulated 
electromagnetic calorimeter {\sf EC2} which is placed symmetrically to 
the beam line and makes angle $-9^{\circ}$ or smaller. \\
\begin{minipage}[c]{1.\linewidth}
\medskip
\psfull
\quad \fbox{
\epsfig{file=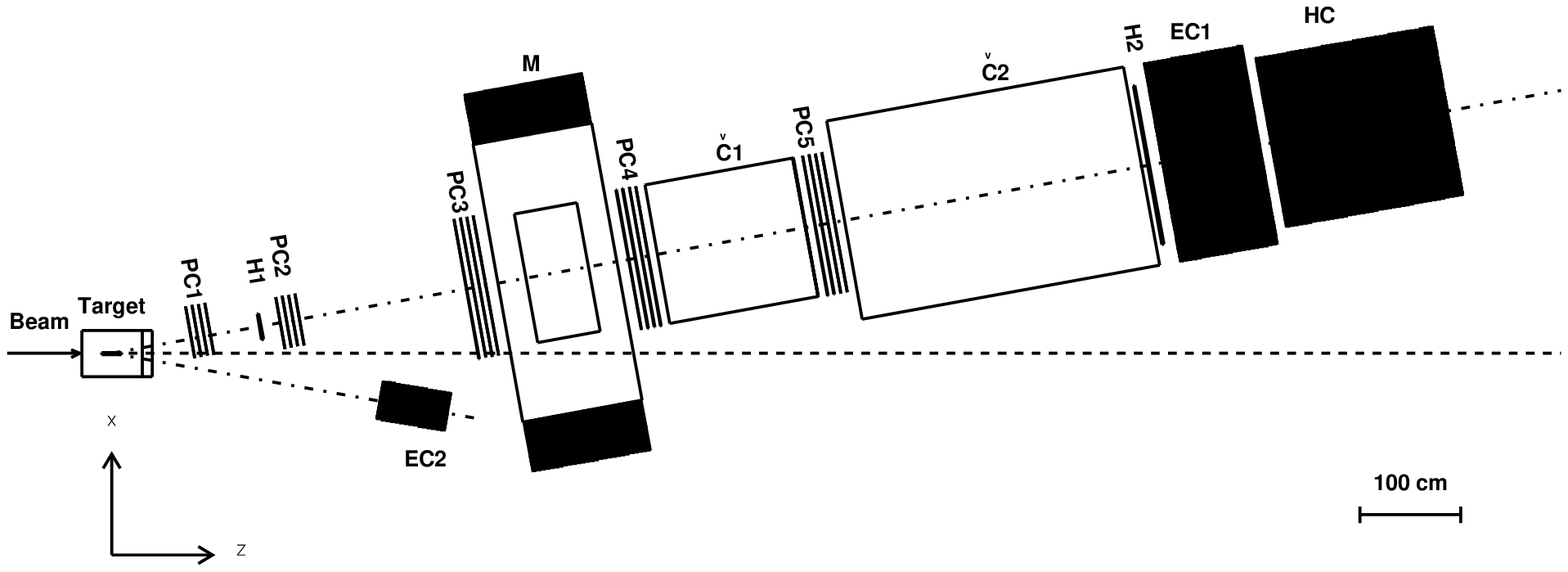,width=12.5cm,%
bbllx=-20pt,bblly=190pt,bburx=550pt,bbury=380pt,clip=}}
{ \footnotesize ~~~~~~~~~~~~~~~~~~~~\\
  Fig. 1. Layout of experimental setup {\sf RAMPEX}:
{\sf PC1}--{\sf PC5} --- blocks of proportional chambers,
{\sf M} --- analysing magnet,
{\sf H1}, {\sf H2} --- trigger hodoscopes,
{\sf \v C1}, {\sf \v C2} --- threshold \v Cerenkov counters,
{\sf EC1}, {\sf EC2} --- electromagnetic calorimeters,
{\sf HC} --- hadron calorimeter.
}
\end{minipage}

 \paragraph  { Beam.} The 70 GeV/c unpolarized proton beam is extracted from 
the accelerator with a bent Si crystal [4], and the measurements with this beam
will take a major part of the experimental program.
The 40 GeV/c $\pi^-$ extracted beam will be also used.
The pion/proton beam intensity is $5 \cdot 10^6$  in a 1-second spill 
with a 9-second interval between spills. 

 \paragraph{ Polarized target.} Propane-diol {\sf $C_3 H_8 O_2$} fills a cavity 
20 mm in diameter and 200 mm in length. The polarization of the hydrogen 
nuclei is about 80\% on average. The dilution factor defined as a ratio of
the number of the target nuclei to the number of polarized nuclei
depends on a type of detected particle and kinematics, and varies between 
about 6 to 10. 
The target contains $9.3 \cdot  10^{24}$ nucleons/cm$^2$. The luminosity 
of the experiment is estimated as 
${\cal L}\sim 5 \cdot 10^{31} {\rm cm}^{-2}{\rm spill}^{-1}$. 

   \paragraph{Magnetic spectrometer.} The magnetic spectrometer 
includes the analysing magnet {\sf M} and 5 blocks of multiwire 
proportional chambers {\sf PC1} --  {\sf PC5}. Each block contains four 
coordinate planes: orthogonal $x$ and $y$ planes and also $u$ and $v$ 
planes which are inclined to $\pm 10^\circ$ with respect to $y$ axis. 
   The magnet {\sf M} has the aperture of $1.3  \times    0.62$~m$^2$
and a length of 0.63 m. The integral of the magnetic field is 
1.0 Tm. The magnet center is placed 4.5 m from the target center.\\
   Two blocks of proportional chambers {\sf PC1} and {\sf PC2} with the
transverse size $530 \times 384$~mm$^2$ are located at distances 0.9 and 
2~m from the target center, respectively. Three blocks of chambers 
{\sf PC3} -- {\sf  PC5} of size $1422   \times   898$~ออ$^2$ are
 placed at the distances 3.7, 5.2 and 7.2 m from the target.\\
The tracking system will allow us to measure charged particle momenta 
with an accuracy $\Delta p/p =  1.7\cdot   10^{-3} p +  2 \cdot  10^{-3}$  
and to reconstruct the straight-line track $x = x_0 + a_x  z$ with an 
accuracy
 $\delta x_0 = 2$~ออ, $\delta a_x = 8.8\cdot 10^{-4}$.

   \paragraph{ Charged particle identification.} Particle types for $\pi$, $K$, 
$p$ and $\bar p$ are determined with the
help of two threshold multi-channel \v Cerenkov counters {\sf \v C1}
and {\sf \v C2} [5].
The counter {\sf \v C1} has 8 channels ($4  \times  2$) and it is filled
with freon-12 at 1 $atm$. The 16-channel counter  {\sf  \v C2}  
($8 \times 2$) is filled with nitrogen also at 1 $atm$.
Combinations of two counters can identify $\pi^{\pm}$ with momenta
$3.1 \div 20$ GeV/c, and $K^{\pm}$ and $p^{\pm}$  from 10 to 20 GeV/c.
   The aperture and the length of \v C1 (\v C2) are  
$1.2 \times 0.9$~m$^2$ and 1.5~m ($1.6 \times 0.88$~m$^2$ and 3.0~m). 

  \paragraph{Electromagnetic (EM) calorimetry.} Two EM calorimeters ({\sf EC1} 
and {\sf EC2} in fig.~1) will detect EM showers. The {\sf EC1} cells are 
made of Pb and scintillator
and have sizes 38$\times$38~mm$^2$ (center) and 76$\times$76~mm$^2$
(periphery). The {\sf EC2} cells of size 38$\times$38~mm$^2$ are made of 
{\sf PbWO}$_4$ [6]. The energy resolutions for {\sf EC1} and 
{\sf EC2} are $ \sigma_E / E \approx 9\% / \sqrt{E} + 0.5\%$ and 
$ \sigma_E /E \approx 3\% / \sqrt{E} + 0.5\% $ respectively.

   \paragraph{Hadron calorimeter.}
  The hadron calorimeter {\sf HC} of compensating type [7] will be used 
to detect $K_L^0$ and neutrons and also as an element of the trigger 
system.
The 10cm$\times$10cm Pb+Sci sandwiches form a matrix of 18$\times$12 
modules (216 channels total) resulting in energy resolution 
 $ \sigma_E /E \approx 57\% /\sqrt{E}$. The ratio of electron
to hadron signals is equal to  $e/h=1.01\pm0.03$. 

 \paragraph{DAQ and trigger.} Data acquisition system and trigger electronics 
are being worked out to fit the data flow of $\sim $2000 events per burst 
with the event size $\sim $ of 1 Mb. The zero level trigger is arranged 
to strobe information from the trigger hodoscopes.The simplest first 
level trigger for the charged arm of the spectrometer is a signal 
coincidence from the two trigger hodoscopes. A special work is made for 
triggering with hadron calorimeter using the proportionality of the 
signal 
$E_x=\sum E_i \cdot sin\theta_{x_i}$ (summation over counters) to the 
transverse momentum $p_T$.
The overall one-charged particle trigger has been worked out
and tested. The multi-particle trigger is in progeress. The details of 
triggering can be found in [8].

\paragraph{First accelerator runs.}
  The {\bf first test run} was performed in Fall'96. The aim was to test 
electromagnetic calorimeter and the related systems of the programming 
shell. 
320 of 1200 {\sf EC1} modules made of Pb+scintillator sandwiches were 
examined and showed the overall energy resolution of 
$\sigma~=~9\% / \sqrt{E} $. 
The {\sf EC1} related data acquisition was carefully tested including 
such elements as read-out electronics, data flow, calibration, LED based
monitoring system and HV power supply. While being 
tested {\sf EC1} moved in two dimensions in the vertical plane with high 
accuracy controlled by a special program. The off-line analysis was also 
undertaken. The results of that testing run are regarded as quite 
satisfactory.\\
The {\bf second test run} has been scheduled for March'97. The program 
includes looking over the remaining 880 {\sf EC1} modules and first 
tests of the tracking system. A certain work will be made to improve 
parameters of the 70-GeV/c proton beam extracted with the bent crystal 
in the 14th channel.\\
The {\bf first data-taking} is being planned for the end of the Fall'97 
run 
supposing detection of $\pi ^o$ and $\eta$ signals. A major part of this
run will be dedicated to the further tests of the tracking system and to
the first launch of the whole setup.
\section{$^3$He target prospects at {\sf RAMPEX}.}
Within any reasonable accelerator run duration, the physical capabilities
of {\sf RAMPEX} is generally restricted by the properties
of the polarized propane-diol target. It becomes opaque at the beam fluxes
more than 10$^7$. So the luminosity $k \cdot 10^{31}$ cm$^{-2}$spill$^{-1}$ is
considered as a maximum. Besides the luminosity related to the collisions 
with the polarized protons in the target is less by a factor of 10.

   Using the internal polarized $^3$He target in the 70-GeV/c proton ring
would provide new possibilities in the
twist-3 and twist-2 studies at {\sf RAMPEX}. In particular, the presence
of polarized neutrons will be very instructive for search of the flavor
dependence of the asymmetry.

  A comparison of polarization effects in inclusive production of particles 
with
various quark content is of special interest.  With the internal $^3$He target 
we could 
compare one-spin asymmetries on polarized protons and polarized neutrons
with appropriate statistics in the following reactions:
\begin{eqnarray}
\nonumber
p + p_\uparrow (n_\uparrow) \rightarrow & \pi^{0} + X    & (d\bar{d})\\
\nonumber       \rightarrow &  K^{0}_{s} + X & (d\bar{s}+\bar{d}s)\\[1.5mm]
\nonumber
p + p_\uparrow (n_\uparrow) \rightarrow & \pi^{+} + X     & (u\bar{d})\\
\nonumber      \rightarrow &  K^{+} + X      & (u\bar{s})\\[1.5mm]
\nonumber
p + p_\uparrow (n_\uparrow) \rightarrow & \pi^{-} + X    & (d\bar{u})\\
\nonumber       \rightarrow &  K^{-} + X     & (s\bar{u}),
\end{eqnarray}
   The most promising are the kinematic regions at $x_F~=~0$ and/or at 
the negative $x_F$.
   The one-spin asymmetry on the
subprocess level which originates from perturbative QCD is proportional to 
the mass of the polarized quark. The masses of $u,d$ quarks are negligibly
small, and it is commonly believed that the strange quarks in polarized proton
 are weakly polarized. 

   On the contrary, the one-spin asymmetry originating from the twist-3
contributions is proportional to the mass parameter $\mu_{{\rm hadr}}$ 
due to the long-distance interactions [9]. As the fragmentation
properties may differ for pions and kaons, the effective size of this
region, $\sim~1/\mu_{{\rm hadr}}$, may also vary, thus resulting in a flavor
dependence of the one-spin asymmetries.

   With the $^3$He target the higher statistics can be achieved at the 
large-$p_T$ twist-3 studies in the processes (1) and the very subtle twist-2
asymmetries in the processes (2). The proton beam intensity in the ring is
equal to $I_{beam}~=~10^{13}$ protons per spill. The length of the spill is 9 
seconds
with the 2-second flat part of the maximum accelerating field in the middle 
of the spill. The frequency of the beam revolutions during this 2-second window
is equal to $n~=~4 \cdot 10^{5}$ spill$^{-1}$. 
Assuming the $^3$He target parameters to be the same as was reported by the 
RIKEN group at SPIN96 in Amsterdam [10], that is the density of 
$5 \cdot 10^{14}$ atoms/cm$^3$ and length of 10 cm, one obtains the target
density of $I_T$ = 1.5$\cdot 10^{16}$ nucleons/cm$^2$.

   This results in luminosity $\cal L$ = $n I_{beam} I_T $ = 
6$\cdot 10^{34}$ cm$^{-2}$s$^{-1}$ which is higher than the current 
{\sf RAMPEX} luminosity by a factor of $10^3$. In a standard one-month
run (2.8$\cdot 10^5$ spills) this corresponds to the statistics of 
10$^4$ events per picobarn. At this level of statistics the twist-3
asymmetry measurements in (1) seem to be reliable up to $p_T~=~$4 GeV/c.
\end{document}